\documentclass[article,twocolumn,amsmath,amssymb,floatfix]{revtex4}
\usepackage{graphicx}
\usepackage{bm}

\newcommand{\be}{\begin{equation}}
\newcommand{\ee}{\end{equation}}
\newcommand{\beq}{\begin{eqnarray}}
\newcommand{\eeq}{\end{eqnarray}}

\begin{document}

\title{Dynamical Quantum Hall Effect in the Parameter Space}

\author{V. Gritsev$^\clubsuit$ and A. Polkovnikov$^\spadesuit$}
\affiliation{$^{\clubsuit}$Physics Department, University of
Fribourg, Chemin du Musee 3, 1700 Fribourg, Switzerland\\
$^{\spadesuit}$ Physics Department, Boston University, Commonwealth
ave. 590, Boston MA, 02215, USA}

\date{\today}

\begin{abstract}

Geometric phases in quantum mechanics play an extraordinary role in
broadening our understanding of fundamental significance of geometry
in nature. One of the best known examples is the Berry
phase (M.V. Berry (1984),  Proc. Royal. Soc. London A, 392:45) which naturally emerges in quantum adiabatic evolution. So far the applicability and measurements of the Berry phase were mostly limited to systems of weakly interacting quasi-particles, where interference experiments are feasible. Here we show how one can go beyond this limitation and observe the Berry curvature and hence the Berry phase in generic systems as a non-adiabatic response of physical observables to the rate of change of an external parameter. These results can be interpreted as a dynamical quantum Hall effect in a parameter space. The conventional quantum Hall effect is a particular example of the general relation if one views the electric field as a rate of change of the vector potential. We illustrate our findings by analyzing the response of interacting spin chains to a rotating magnetic field. We observe the quantization of this response, which term the rotational quantum Hall effect.

\end{abstract}

\maketitle

\section{Introduction}

In quantum mechanics the Berry phase is defined as a phase accumulated
by the wave function during the adiabatic evolution around a closed
path in the parameter space denoted by $\vec s$. This phase can be
obtained by integrating the Berry connection:
\be
\mathcal A_\mu=i\langle \psi|\partial_\mu|\psi\rangle
\ee
along this path~\cite{berry_84}. Here to shorten notations we define
$\partial_\mu\equiv \partial_{s_\mu}$. The vector $\vec s$ stands
for an arbitrary set of parameters which change during the adiabatic
evolution. These can be a coordinate of a particle, its momentum, magnetic or electric field, vector potential, pressure, volume, a coupling to some external potential and so force. From the definition above it is clear that the Berry phase has a purely geometric interpretation. Together with the Berry connection $\mathcal A_{\mu}$ (analogous to the vector potential in the parameter space) one defines the Berry curvature (analogous to the magnetic field):
\be
\mathcal F_{\mu\lambda}=\partial_\mu \mathcal A_{\lambda}-\partial_\lambda \mathcal A_{\mu}.
\ee
From the Stokes theorem it follows that the Berry phase along some
closed path can be found by integrating the Berry curvature over the
area enclosed by this path.  A non-trivial Berry phase, i.e. the phase not equal to $0$ or $\pi$ and a non-zero Berry curvature are in general associated with a broken time-reversal symmetry, because otherwise all wave functions can be made real~\cite{griffiths}. 

The Berry curvature is directly related to the geometric
tensor~\cite{provost_80} introduced to describe manifolds of
adiabatically connected wave-functions $\psi(\vec s)$:
\be
\chi_{\mu\lambda}=\langle\psi|\overleftarrow{\partial_\mu}\partial_\lambda|\psi\rangle-
\langle \psi|\overleftarrow{\partial_\mu}|\psi\rangle\langle\psi|\partial_\lambda|\psi\rangle
\label{geom:tens}
\ee
It is straightforward to check that $\mathcal F_{\mu\lambda}=-2\Im
[\chi_{\mu\lambda}]$, while $g_{\mu\lambda}=\Re [\chi_{\mu\lambda}]$
defines the Riemannian metric tensor associated with the same
manifold. The Riemann curvature associated with the above metrics as
well as the components of the geometric tensor can serve as
observable independent measures of singularities like phase
transitions~\cite{venuti_07}. The geometric tensor can be extended
to mixed states by taking the statistical average of the geometric
tensor associated with individual pure states .

The Berry phase and related concepts found multiple applications in many
different quantum and classical systems
In particular, it underlies the Aharonov-Bohm effect; it shows up in
transport in graphene in a quantum Hall regime~\cite{zhang_2005}; it can emerge in 
photon interference of a circularly polarized light~\cite{tomita_86,
haldane_86}; it appears in anomalous quantum Hall effect in magnetic
metals~\cite{haldane_04}, in various magnetoelectric response functions~\cite{ortiz_book},  in Thouless pumps~\cite{thouless_83, thouless_84} and many other phenomena, including various forms of solid-state related topological phenomena like topological
insulators (see Ref.~\cite{niu_10} for a recent review). Most known applications of the Berry phase and ways to
experimentally measure it rely on existence of free or nearly free
quasi-particles which can independently interfere affecting
transport properties. If the concept of quasi-particles is ill
defined and transport experiments are not feasible, like e.g. in
cold atom systems, the Berry phase could not be measured using
traditional approaches.

\section{Results}

The main finding of our paper is that in slowly driven isolated systems the Berry curvature emerges in a linear response of physical observables to the quench velocity $\vec v=\dot{\vec s}$.  Specifically we show that if the quench velocity $\vec v$ is aligned along the $\lambda$ direction in the parameter space then measuring the response of the generalized force along the $\mu$-direction: $\mathcal M_\mu=-\langle \psi(t_f)| \partial_\mu
\mathcal H| \psi(t_f)\rangle$ gives the $\lambda\mu$ component of
the Berry curvature:
\be
\mathcal M_\mu={\rm const}+\mathcal F_{\mu\lambda}v_{\lambda} +\mathcal O(v^2).
\label{eq:main}
\ee
where we used the Einstein convention of summation over repeated
indexes. The constant term gives the value of the generalized force
in the adiabatic limit. This result is very remarkable since it shows that even without dissipation the leading non-adiabatic response of a quantum system is local in time. Indeed it is determined by the instantaneous velocity and the instantaneous matrix elements and the instantaneous spectrum. We sketch the details of the derivation of Eq.~(\ref{eq:main}) in the Section Methods. Here we only point that this result is valid if either of the three conditions are met (i) the velocity $\vec v$ is turned on smoothly, (ii) the system is prepared initially in a state with a large gap, (iii) there is a weak dephasing mechanism in the system and the time of experiment is longer than the dephasing time. The first two conditions imply that the system is not excited at the initial time of the evolution and the last condition implies that even if there are small excitations in the system, they come with a random phase. 
Note that the dephasing does not have to be due to external noise, it can be e.g. due to averaging of over different experimental runs with slightly fluctuating durations. The equation (\ref{eq:main}) applies to both gapless and gapped systems at either zero or at finite temperatures. It can be used for a single particle and for an interacting many-body system in a thermodynamic limit. However, we need to keep in mind that the Berry curvature is a susceptibility, in particular it can be expressed through the non-equal time correlation functions:~\cite{berry_84}
\be
\mathcal F_{\mu\lambda}=-i\int_0^\infty dt\, t \langle [\partial_\mu \mathcal H(t),\partial_\lambda \mathcal H(0)]\rangle_{0},
\ee
Here $\partial_\mu \mathcal H(t)$ stands for the operator $\partial_\mu \mathcal H$ in the Heisenberg representation with respect to the instantaneous Hamiltonian, $[\dots]$ stands for the commutator and the subindex $0$ means that the average is taken with respect to the adiabatically evolved state (e.g. the instantaneous ground state if the system is initially prepared at zero temperature). As usually we assume that the energies have infinitesimal positive imaginary parts to guarantee the convergence of the integral above.  In low dimensional gapless systems or near phase transitions the Berry curvature can diverge. Then the linear response theory breaks down and the dependence $\mathcal M_\mu(v_\lambda)$ can become non-analytic. In this work we will not be concerned with these quite special situations. Let us point that in large systems if we are dealing with extensive couplings the Berry curvature is also extensive. Therefore possible non-extensive number of degeneracies of the ground state does not affect the result Eq.~(\ref{eq:main}). However, these degeneracies can affect topological protection of the integer Chern numbers associated with the Berry curvature which we discuss below. Let us also point that Eq.~(\ref{eq:main}) implies that the non-vanishing linear response coefficient indicates a broken time reversal symmetry in the Hamiltonian (possibly by the coupling $s_\mu$). This situation is opposite to that in the imaginary time dynamics, where the linear response for a similar protocol reveals the symmetric part of the geometric tensor~\cite{degrandi_11}. The latter is nonsensitive to
the time-reversal symmetry. 

The equation~(\ref{eq:main}) can be interpreted as a Hall effect in the abstract parameter space. Indeed then $v_\lambda$ can be interpreted as a driving current in $\lambda$-direction. The response ${\cal M}_\mu$ is analogous to the electric field in the transverse direction and then $\mathcal F_{\lambda\mu}$ is the analogue of the Hall resistance. Conversely by $s_\lambda$ we can understand the electric field component in the $x$-direction (which can be viewed as a rate of change of the x-component of the vector potential). Similarly by $s_\mu$ we can understand the vector potential along $y$-direction. Then the corresponding generalized force $\mathcal M_\mu=-\partial_\mu \mathcal H$ is proportional to the $y$-component of the current. Then up to the coupling $e^2/h$ the Berry curvature is given by the Hall conductivity and Eq.~(\ref{eq:main}) reduces to the well known result (see Sec.~\ref{quantum_hall} for more details). Different variations of Eq.~(\ref{eq:main}) also appeared earlier in other particular contexts~\cite{thouless_84, berry_robbins}

The proposed method of finding the Berry curvature has another significant advantage over traditional interference based methods. Namely, it does not require strict adiabaticity, which is nearly impossible to achieve in large interacting systems. Contrary, the applicability of
the linear response regime, where Eq.~(\ref{eq:main}) is valid, only
requires that intensive quantities like the magnetization per unit
volume remain small. Therefore there are no difficulties related to
taking the thermodynamic limit with possible exceptions near
singularities like critical points where the Berry curvature can
diverge. The relation (\ref{eq:main}) gives a clear route
for measuring the Berry curvature and consequently the Berry
connection and the Berry phase in generic systems. A possible
procedure can consist of evolving the system initially prepared in
the ground state by quenching some tuning parameter $s_\lambda$
smoothly in time.  Then at the time $t=0$ corresponding to the Hamiltonian of interest one measures the generalized force corresponding to a different parameter $s_\mu$: $\mathcal M_\mu$.
Repeating these measurements at different velocities one can extract
the slope of $\mathcal M_\mu(v_\lambda)$, which coincides with the
Berry curvature. One can repeat the same sequence at different
values of $\vec s$ confined within some closed area of the parameter
space and evaluate the Berry phase using the Stokes theorem.
The relation~(\ref{eq:main}) also extends to weak continuous
measurements. Then the Berry curvature can be extracted from the
linear in velocity correction to $\mathcal M_\mu$.

If the parameter field $\vec s$ lies on an arbitrary compact
manifold (surface) ${\cal S}$ like a sphere or a torus then the integral over the Berry curvature for a given state $|\psi\rangle$ forms a topological integer invariant known as the first Chern number:
\be
ch_{1}(|\psi\rangle)=\frac{1}{2\pi}\int_{{\cal S}}d S_{\mu\lambda} \mathcal F_{\mu\lambda}.
\label{eq:chern}
\ee
where $dS_{\mu\lambda}$ is an area form element in the parameter space. To observe the quantization It is important that the
gap separating the ground state from the rest of excitations never closes on this surface. The regions in the parameter space, where there are such nodes on the surface define the crossover between different quantized plateaus. Thus Eq.~(\ref{eq:main}) allows one to experimentally or numerically map the manifold of degeneracies in the ground state wave function. We will illustrate this point below using an example of an anisotropic spin chain. If in addition the surface can be represented as an invariant closed manifold (such that $\mathcal F_{\mu\lambda}$ is constant on this manifold) then Eq.~(\ref{eq:main}) gives a generalization of the integer quantum Hall effect to the abstract parameter space with $\mathcal F_{\mu\lambda}$ being a product of a factor of $2\pi$ and
an integer divided by the surface area. This setup becomes analogous
to the Thouless pump~\cite{thouless_83, thouless_84} with the
difference that the quench parameter and the observable are
arbitrary not necessarily related to the particle transport.

\section{Examples}

\begin{figure}[ht]
\begin{center}
\includegraphics[width=8cm]{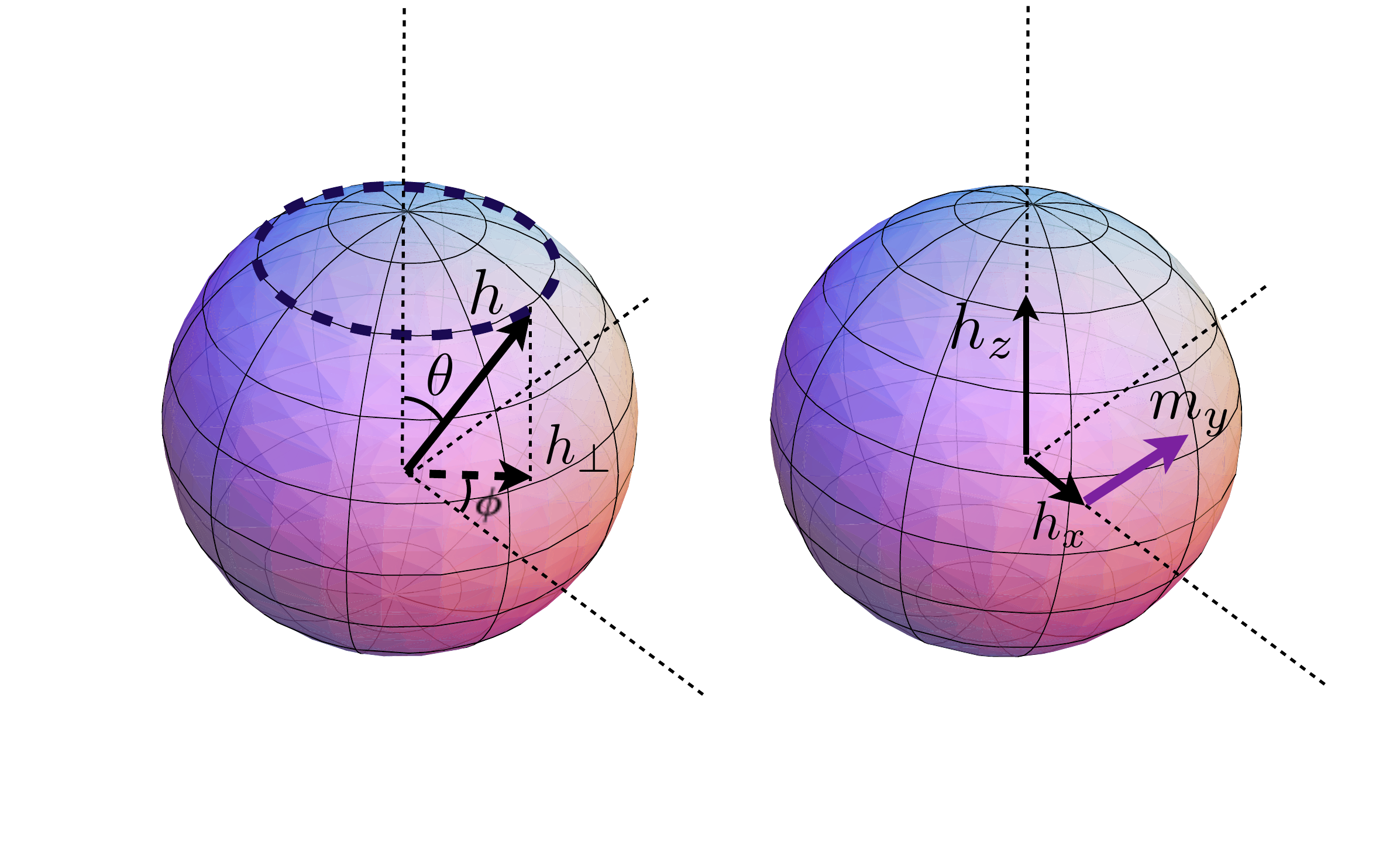}
\vspace{-1cm}
\caption{
Two ways of measuring the Berry phase for a spin in a magnetic field:
 (i) Traditional (left panel) where an external parameter adiabatically changes
 along a closed path, e.g. by varying the angle $\phi$. Then the Berry phase can
 be measured through the interference signal between the original and the
 adiabatically evolved spins. (ii) Dynamical (right panel). In this setup one
 changes an external parameter $h_x$ linearly in time with some velocity $v_x$ and measures the response of the magnetization $m_y(v_x)$. The linear slope of the latter gives the Berry curvature $\mathcal F_{yx}$ (see Eq.~(\ref{eq:main})).}
\label{fig:spin}
\end{center}
\end{figure}

To illustrate how this idea works we will use several specific 
examples gradually increasing their complexity. First let us consider a spin one half particle in the external magnetic field described by the Hamiltonian:
\be
\mathcal H=-\vec h\cdot \vec\sigma,
\ee
where $\vec \sigma$ stands for Pauli matrices. It is well
known~\cite{sakurai} that if we choose a path where $h_z$ is
constant, $h_x=h_\perp \cos\phi, \; h_y=h_\perp\sin\phi$ and the
angle $\phi$ changes by $2\pi$ (see Fig.~\ref{fig:spin}) then the
ground state of the spin acquires the Berry phase:
\be
\gamma=\pi (1-\cos(\theta))=\pi\left(1-{h_z\over h}\right)
\ee

A conventional way of measuring the Berry phase in this setup will
involve adiabatic motion of the spin around this path and looking
into a signal sensitive to the interference of the rotated and the
original spins.

\begin{figure}[ht]
\begin{center}
 \includegraphics[width=8.5cm]{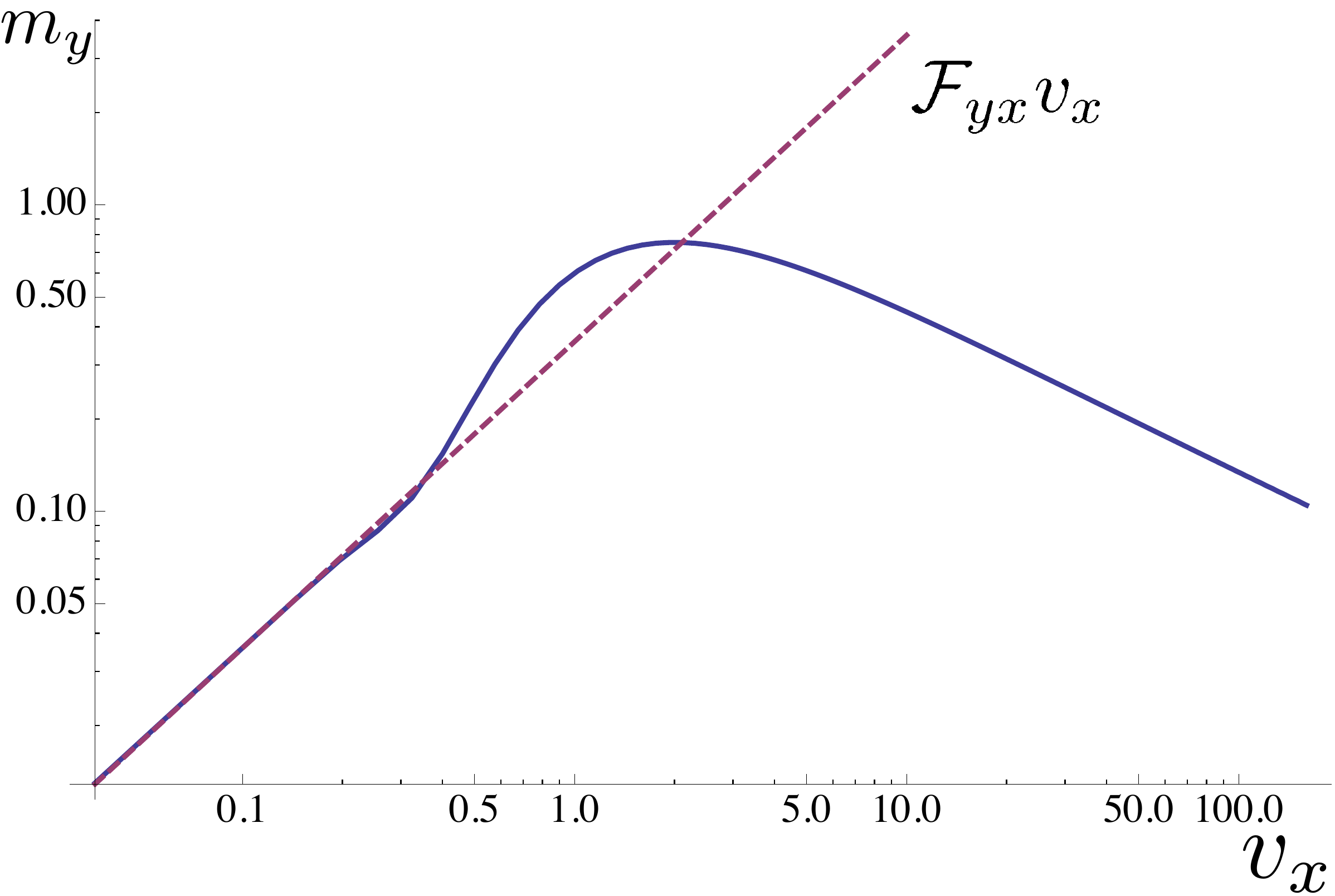}
\caption{
Magnetization along the $y$-axis as a function of the quench
velocity of the $x$-component of the magnetic field evaluated at $h_x=0.5$, $h_y=0$ and $h_z=1$. The dashed line is the linear response prediction given by Eq.~\ref{eq:main}.}
\label{fig:m_y}
 \end{center}
\end{figure}

Using Eq.~(\ref{eq:main}) one can obtain the same result in a
non-equilibrium protocol which does not directly involve any
interference. We will illustrate this with a specific setup. First
we prepare the system in the ground state of a magnetic field having
a large negative $x$-component and fixed $z$-component at the moment
$t=t_0$. Then we evolve the magnetic field along the $x$-direction
linearly in time: $h_x(t)=h_x+v_x t$. At time $t=0$ we measure the
magnetization along the $y$-direction: $m_y=\langle
\psi(0)|\sigma_y|\psi(0)\rangle$. This problem is simple enough so
that it can be solved analytically using the Weber
functions~\cite{vitanov_99}, but the solution is quite involved so
we solve the Schr\"odinger equation numerically instead. The result
of these simulations for $h_x(t_0)=-99.5$, $h_x=h_x(t=0)=0.5$ and
$h_z=1$ is shown in Fig.~\ref{fig:m_y} (solid line).
The Berry curvature for this system can be easily computed using
e.g. explicit form of the ground state wave function and
Eq.~(\ref{geom:tens}). In particular, for $h_y=0$ it reads:
\be
\mathcal F_{yx}={h_z\over 2 h^3}={h_z\over 2\left(h_x^2+h_z^2\right)^{3/2}}
\ee
The linear function $ \mathcal F_{yx} v_x$ (dashed line) is in perfect
agreement with the low velocity asymptotic of the exact solution.
From this Berry curvature we can immediately infer the Berry phase
along the circular path by noting the rotational invariance of the
system:
\be
\gamma=\int\limits_0^{2\pi} d\phi\int\limits_0^{h_\perp} d\eta\, \eta {h_z\over 2(h_z^2+\eta^2)^{3/2}}
=\pi\left(1-{h_z\over h}\right),
\ee
which is of course the correct result. If the rotational symmetry is
broken, one would need to evaluate $\mathcal F_{yx}$ in a
sufficiently dense set of parameters $h_x$ and $h_y$ enclosed by the
closed path and evaluate the area integral over the Berry
curvature using finite differences.

To get the first Chern number in this example we can use $h_\phi$
and $h_\theta$ as the external parameters keeping the total magnetic
field fixed. Then a similar quench procedure  will result in
$m_\phi\approx \mathcal F_{\phi\theta} v_\theta$, where $\mathcal
F_{\phi\theta}=1/2\sin(\theta)$, which after integration over the
spherical surface will result in
\be
ch_1={1\over 2\pi}\int_0^{2\pi} d\phi \int_0^\pi d\theta \,{1\over 2}\sin(\theta) =1.
\ee
The quantization unit here  is the system's spin $s=1/2$.

Next we extend the example above to a
Heisenberg spin chain in a magnetic field:
\be
\mathcal H=-\sum_{j=1}^N {\vec h}{\vec \sigma_j}-J\sum_{j=1}^{N-1} {\vec \sigma_j}{\vec\sigma_{j+1}},
\label{h_spin_chain}
\ee
where $N$ is the chain size. In this setup we will analyze the response to rotating magnetic
field in the $\theta$ direction (see Fig.~\ref{fig:spin}) fixing the
magnitude $h=1$:
\be
h_x(t)=\sin\left({v^2 t^2\over 2\pi}\right), \; h_z(t)=\cos\left({v^2 t^2\over 2\pi}\right),\; h_y(t)=0.
\label{ht}
\ee
This choice of time dependence guarantees that the angular velocity is turned on smoothly and the system is not excited at the initial moment of evolution. At the point of measurement $t=\pi/v$ the velocity of the $\theta$-component of the magnetic field is exactly $v$. As in the previous example we numerically solve the time dependent Schr\"odinger equation and evaluate the magnetization along the $y$ axis: $m_{\phi}=\sum_j \langle \sigma_j^y\rangle$ as a function of the velocity. The slope $\mathcal F_{\phi\theta}$ must give the Berry curvature, which we compute as a function of the interaction coupling $J$.

In order to check the quantization of the first Chern number in this
system one needs to integrate the extracted value of $\mathcal
F_{\phi\theta}$ over the sphere $h=1$. However, in this case it is not necessary because of the rotational invariance of the interaction. Thus the integration will result in just the multiplication of $\mathcal F_{\phi\theta}$ by the area of the sphere, which is $4\pi$. We thus anticipate that $\mathcal F_{\phi\theta}$ is quantized in units of $1/2$ irrespective of the interaction coupling.
\begin{figure}[t]
\begin{center}
 \includegraphics[width=9.2cm]{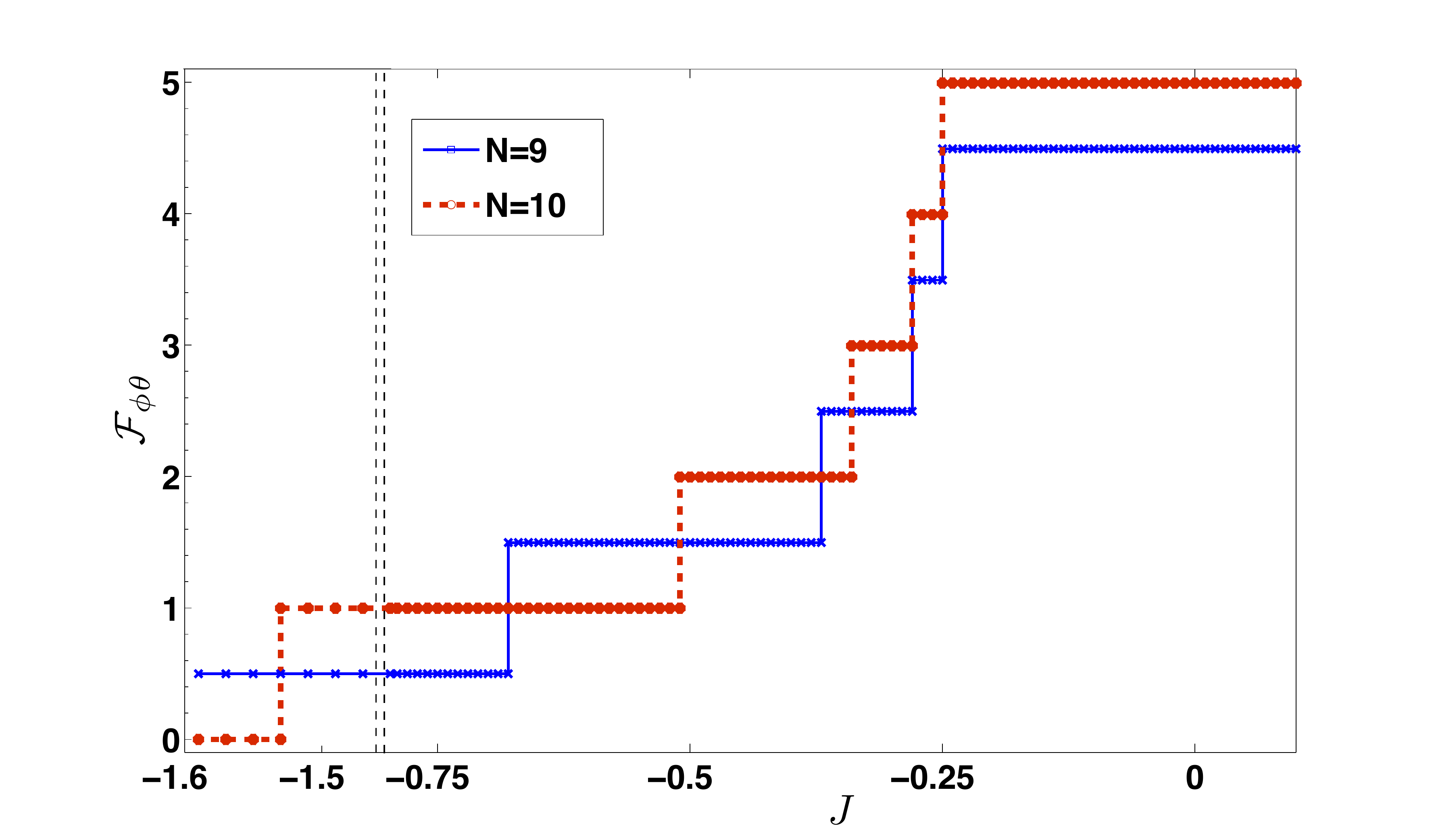}
\caption{ The Berry curvature extracted from numerically evaluating
dynamical response of the magnetization of a Heisenberg spin chain
to the rotating magnetic field. The two lines represent chains of
the length $N=9$ and $N=10$. The quantization plateaus clearly
indicate the topological character of the response. Numerical
simulations were done by solving time dependent Schr\"odinger
equation with fixed velocity $v=0.1$ (see Eqs.~(\ref{h_spin_chain})
and (\ref{ht})), The Berry curvature was extracted from the
transverse magnetization: $\mathcal F_{\phi\theta}\approx
m_y(v)/v$.}
\label{fig:spin_chain}
 \end{center}
\end{figure}
This is indeed what we observe (Fig.~\ref{fig:spin_chain}).  In
the ferromagnetic regime ($J\geq 0$) the Berry curvature is equal to
$N/2$, which indicates that the ground state behaves as a collective
spin of magnitude $S=N/2$. In the antiferromagnetic regime $J$ is large
and negative such that the system behaves as a spin singlet $S=0$ for even $N$
and as an effective spin $1/2$ for odd $N$.  The transition between the spin singlet (for even chain) and the maximally polarized state occurs through the quantized steps, which reflect the total value of the spin in the initial state. It is interesting to note that even though we numerically extracted the slope of magnetization $\mathcal F_{\phi\theta}\approx M_y(v)/v$ from a moderately small velocity $v=0.1$ the accuracy of the quantization of plateaus is better than $0.1\%$. 

The example above has still one significant simplification coming from the fact that the magnetization commutes with the Heisenberg interaction term. Therefore the time evolution of the magnetization is decoupled from the latter. To show that the quantization of the dynamical response holds in a more generic setup  we will next consider a disordered (and hence nonintegrable) Heisenberg chain with the Hamiltonian
\be
\mathcal H=-{\vec h}\sum_{j=1}^N \xi_j {\vec \sigma_j}-J\sum_{j=1}^{N-1} \eta_j{\vec \sigma_j}{\vec\sigma_{j+1}},
\label{h_spin_chain_1}
\ee
where $\xi_j$ and $\eta_j$ are random variables, which for concreteness are both chosen from a box distribution in the interval $[0.75,1.25]$.
\begin{figure}[t]
\begin{center}
 \includegraphics[width=9.2cm]{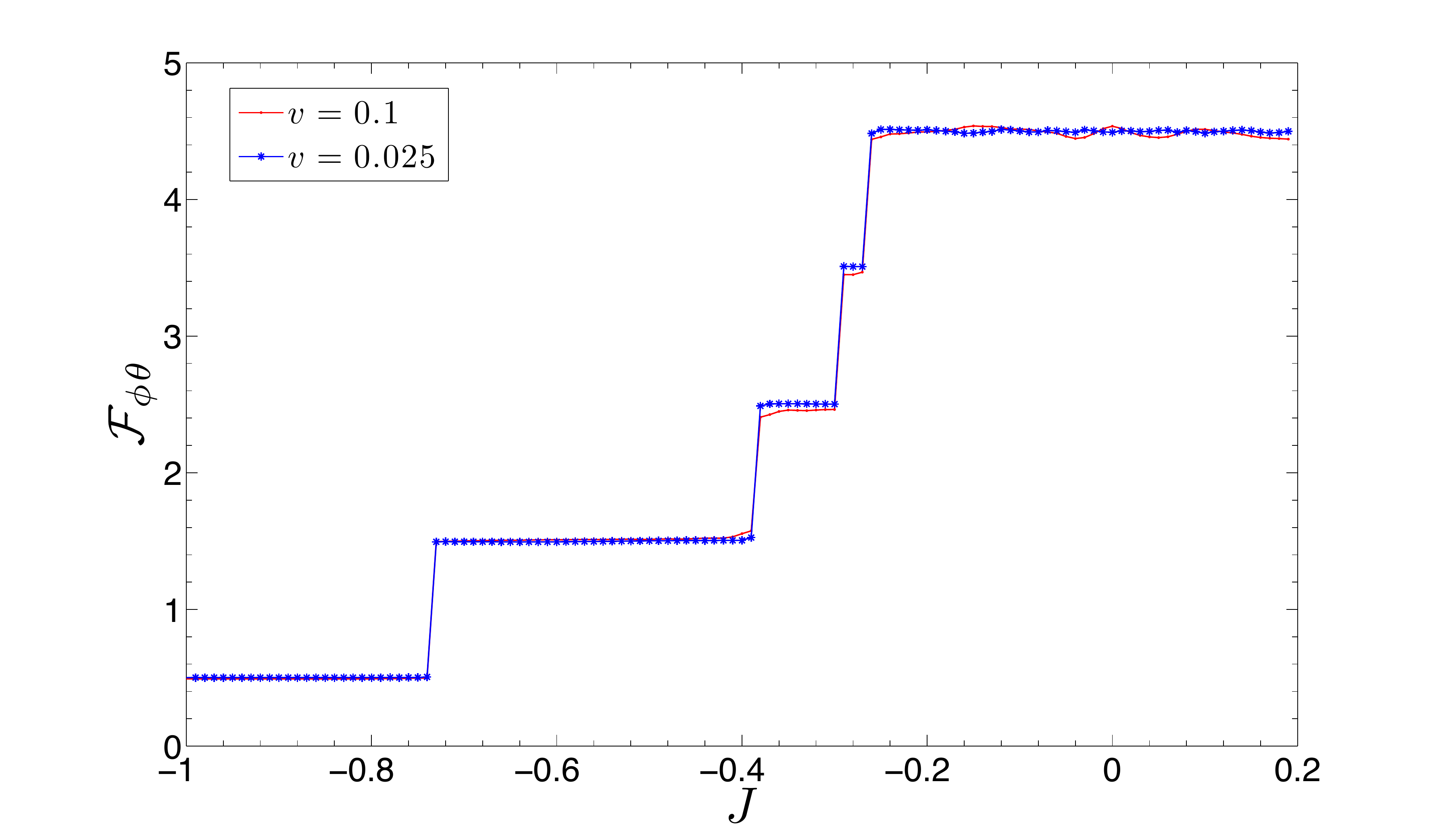}
\caption{ Berry curvature extracted from dynamical simulations of the disordered Heisenberg chain (see Eq.~(\ref{h_spin_chain_1})). The simulations are done for the spin chain of length $N=9$ using the protocol identical to that in Fig.~\ref{fig:spin_chain} for a particular realization of the disorder. The two lines represent two different velocities $v=0.1$ and $v=0.025$. Smaller velocity clearly improves the accuracy of quantization.}
\label{fig:spin_chain_1}
 \end{center}
\end{figure}
We repeat the same protocol as before, i.e. change the magnetic field according to Eq.~(\ref{ht}), and extract the Berry curvature from  the response of the magnetization to the velocity $v$. In Fig.~\ref{fig:spin_chain_1} we show the results of such simulations for a chain of length $N=9$ with a given realization of disorder. The slopes are extracted from two different velocities $v=0.1$ and $v=0.025$. The plot clearly shows that the quantization of the response persists. At higher velocity the crossovers between the plateaus are slightly more rounded and one observes small fluctuations of the numerically extracted $\mathcal F_{\phi\theta}$ in the plateau regions. At smaller velocity, i.e. closer to the linear response regime, these fluctuations are suppressed and we see again nearly perfect quantization.

As a final example we analyze an anisotropic chain:
\be
\mathcal H=-{\vec h}\sum_{j=1}^N {\vec \sigma_j}-J\sum_{j=1}^{N-1} (\vec{\sigma}_j^{\perp}\vec{\sigma}_{j+1}^{\perp}+0.75\sigma_j^z\sigma_{j+1}^z),
\label{h_anis}
\ee
where $\vec{\sigma}_j^{\perp}$ denotes $x,y$ components of the spins. This chain has only azimuthal symmetry in the $xy$ plane. In order to get the quantization of the response we thus need to average the Berry curvature over the polar angle of the magnetic field with respect to the $z$-axis. If we are taking a weak continuous measurement then this average is equivalent to the time average.
\begin{figure}[t]
\begin{center}
 \includegraphics[width=9.2cm]{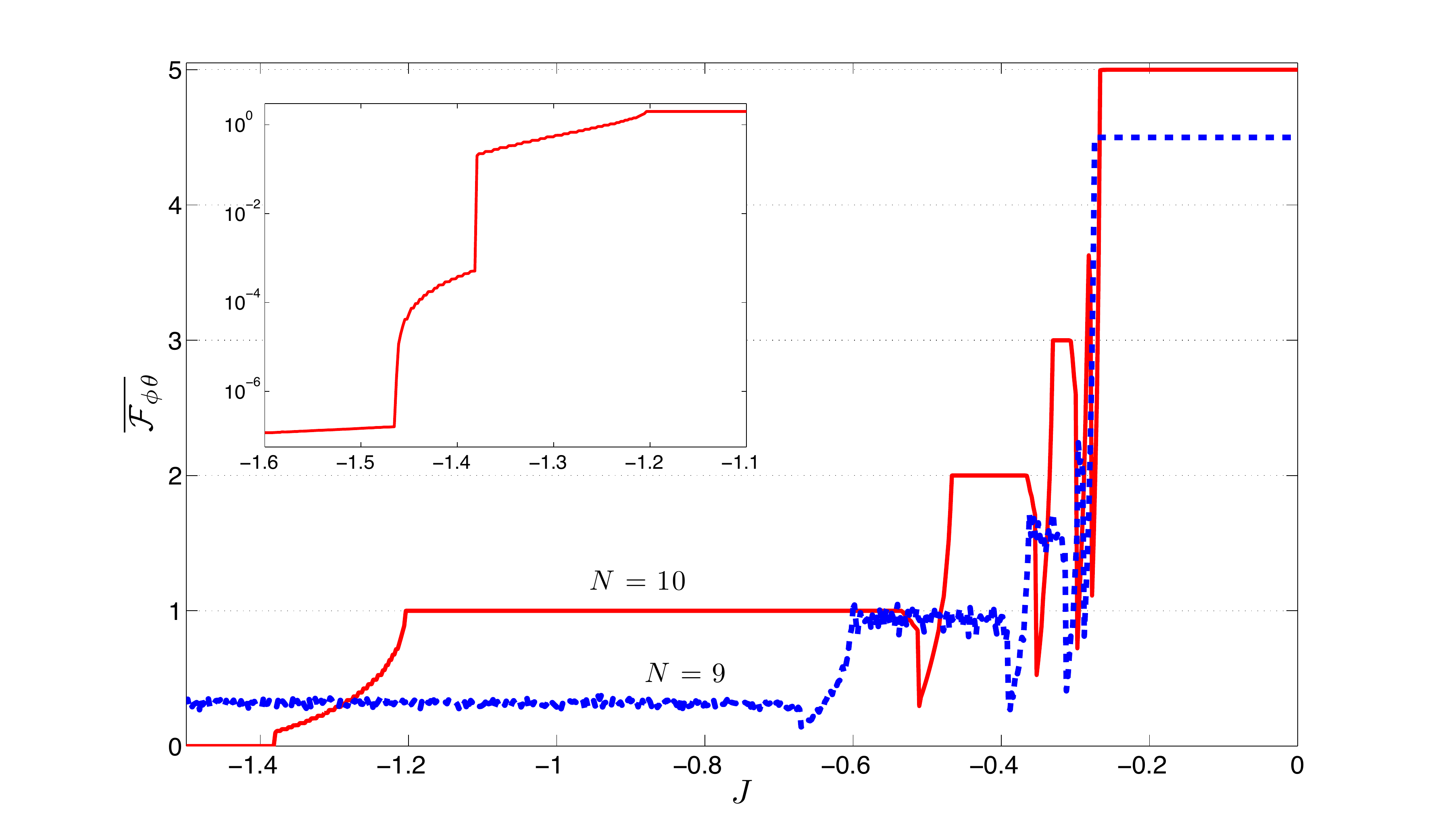}
\caption{Berry curvature averaged over the polar angle for the anisotropic spin chain chain (see Eq.~(\ref{h_anis})). The simulations are done for spin chains of lengths $N=9$ and $N=10$ and the angular velocity $v=0.025$. The inset shows the result for the even chain in the log scale. Both even and odd chain show a clearly quantized plateau at $J\gtrsim -0.25$ at the value $N/2$. The even chain also shows well defined plateaus at other integer values. The regions between the plateaus correspond to the regions where the degeneracies of the ground state cross the surface $h=$const.}
\label{fig:spin_chain_2}
 \end{center}
\end{figure}
The results of the numerical simulations for even and odd chains with lengths $N=9,10$ are plotted in Fig.~\ref{fig:spin_chain_2}. To obtain these results we used the same protocol of changing the magnetic field as in Eq.~(\ref{ht}) except that we performed an additional averaging over the polar angle. Both chains show clear quantization of the response with $\overline F_{\phi\theta}=N/2$ for $J\gtrsim -0.25$. But at smaller values of $J$ the behavior of the response is different. For odd chain the plateaus have very large fluctuations while for the even chain the plateaus are perfectly defined. We can understand the odd chain result at e.g. large negative $J$ as coming from the double degeneracy of the ground state where the unpaired spin can be localized on the left or right edges of the chain (or alternatively degeneracy between symmetric and antisymmetric combinations of these spins). In a rotationally invariant system these states are protected by the symmetry and 
there are no transitions between them so the system behaves as effectively gapped. In an anisotropic chain this protection is lost and thus the ground state is not protected, hence we see no accurate quantization of the plateaus. For the even chain the intervals between the plateaus correspond to the regions where the ground state degeneracies cross the integration surface of constant magnetic field $h=1$. By deforming the shape of the integration surface such that the jumps between the plateaus become sharp one can map the precise location of the degeneracies. Analysis of the microscopic origin of this response as well as understanding which plateaus (except for the trivial ones with $\overline F_{\phi\theta}/N=0, 1/2$) survive in the thermodynamic limit is beyond the scope of this work. 

In the examples above we focused on magnetic systems. But our analysis goes through if we consider e.g. dipoles in a time-dependent electric field or other setups. For isotropic systems the quantized Berry curvature reveals the total spin of the system. On a more fundamental level this quantization reveals the number of non-equivalent submanifolds of degeneracies of the ground state within the integration surface. In the analyzed examples for illustration purposes we assumed that the $g$-factor, i.e. the coupling of spin to the magnetic field, is unity. In real experiments this quantization of the response can be used for e.g. a precision measurement of the $g$-factor like the ordinary quantum Hall effect is used for the precision measurement of $e^2/h$. Quantization of the integrated Berry curvature can also be used for canceling effects of random static magnetic fields, which might affect accuracy of direct measurements of the magnetization. These and other potential applications will be a subject of future investigation.

\section{Discussion and Conclusions}. 

Our approach allows an interesting
possibility of mapping the Hilbert space  topology
through quantum dynamics.
By measuring the Berry curvature one can experimentally analyze topological properties of the ground states, extract information about their possible degeneracies and analyze phase transitions between different topological states. One can also use 
Berry curvature as a probe of time reversal symmetry breaking in complicated systems e.g. in biology.  

In conclusion we demonstrated that the Berry curvature can be
measured in generic systems, interacting or not, as a leading
non-adiabatic response of physical observables to quench velocity.
This method does not require stringent adiabatic conditions hard to
achieve in large systems. While in this paper we focused on the quantum dynamics close to the ground state, our main result Eq.~(\ref{eq:main}) applies to the mixed states as well. We illustrated applicability of this
method by numerically solving the time dependent Schr\"odinger
equation for a single spin and different interacting spin chains. In all the cases we found quantization of the dynamical response in agreement with theoretical expectations. This quantization can be interpreted as a dynamical quantum Hall effect. Our findings also reveal deep connections between quantum dynamics and equilibrium geometric properties associated with the adiabatically connected wave functions. We believe that these
findings can be used in a variety of systems to experimentally
analyze the valuable geometrical properties of the interacting
systems without need to perform sensitive and not always feasible
interference experiments. 

\section{Methods.}

\subsection{Sketch of the derivation of Eq.~(\ref{eq:main})}

Derivation of Eq.~(\ref{eq:main}) is rather simple. It relies on the
adiabatic perturbation theory, i.e. perturbation theory in the
instantaneous basis~\cite{ortiz_2008, degrandi_2009}. For linear
quenches one finds that the transition amplitude to the $n$-th
eigenstate of the final Hamiltonian is~\cite{degrandi_2009}
\be
a_n\approx i v_\lambda {\langle n|\partial_\lambda|0\rangle\over (\mathcal
E_n-\mathcal E_0)}\mathrm
e^{-i\Theta_{n0}}\biggl|_{\lambda_i}^{\lambda_f}=- i v_\lambda {\langle
n|\partial_\lambda \mathcal H|0\rangle\over (\mathcal E_n-\mathcal
E_0)^2}\mathrm e^{-i\Theta_{n0}}\biggl|_{\lambda_i}^{\lambda_f},
\label{tr_amp}
\ee
where $\Theta_{n0}$ is the full phase difference (including the
dynamical and the Berry phases) between the $n$-th and the ground
instantaneous eigenstates during time evolution:
\be
\Theta_{n0}(\lambda)=\int\limits_\lambda^{\lambda_f} d\lambda'
\left[{\mathcal E_n(\lambda')-\mathcal E_0(\lambda')\over
v(\lambda')}-i(\mathcal A_n(\lambda')-\mathcal
A_0(\lambda'))\right].
\ee

If the initial state has a large gap or if the protocol is designed in
such a way that the initial evolution is adiabatic then
Eq.~(\ref{tr_amp}) takes a particularly simple form:
\be
a_n\approx -i v_\lambda {\langle n|\partial_\lambda \mathcal H|0\rangle\over (\mathcal E_n-\mathcal E_0)^2}\biggr|_{\lambda_f}
\label{tr_amp1}
\ee
The contribution of the initial term in Eq.~(\ref{tr_amp}) to the
expectation value of the off-diagonal observables can be
additionally suppressed by the fast oscillating phase $\Theta_{n0}$.
This suppression will happen even if by magnitude this contribution
is comparable to the final term~(\ref{tr_amp1}).  Let us note that
in order to obtain this result from Eq.~(19) in
Ref.~\cite{degrandi_2009} one needs to perform an additional phase
transformation to undo the transformations given by Eqs. (6) and
(14) of that work. Let us also point that Eq.~(\ref{tr_amp1}) is equivalent to Eq. (2.10) in Ref.~\cite{thouless_84} for a particular choice of parameters. However, as we discuss above this result is only valid provided that the initial term in a more general Eq.~(\ref{tr_amp}) is unimportant.

From Eq.~(\ref{tr_amp1}) it is straightforward to derive Eq.~(\ref{eq:main}):
\beq
&&\mathcal M_\mu=\langle \psi|-\partial_\mu \mathcal H|\psi\rangle\approx
\langle 0|-\partial_\mu \mathcal H|0\rangle\nonumber\\
&&+iv_\lambda  \sum_{n\neq 0}
{\langle 0|\partial_\mu \mathcal H|n\rangle\langle n|\partial_\lambda \mathcal H|0\rangle-\mu\leftrightarrow\lambda \over (\mathcal E_n-\mathcal E_0)^2}
\eeq
It is easy to check that the second term in this equation is equivalent to Eq.~(\ref{eq:main}).

\subsection{Application of Eq.~(\ref{eq:main}) to the integer quantum Hall effect.}
\label{quantum_hall}

He we elaborate on that the quantum Hall effect can be
understood as a particular application of Eq.~(\ref{eq:main}). This discussion will closely follow the first chapter of Ref.~\cite{stone_92}.

The Hall current in the ordinary setup can be derived using the
adiabatic transport theory combined with the Kubo formula. To
compute the Hall conductivity one usually applies the adiabatic
formulas similar to those derived in Appendix A, to some in general
interacting Hamiltonian $\mathcal H(A_{x}(t),A_{y}(t))$ on a {\it
torus} of size $L_{x}\times L_{y}$. The Hamiltonian $\mathcal H$
depends on the vector potential of the fixed external magnetic field
$A^{mag}$ and on slowly varying perturbations $\theta$ and $\phi$,
\beq
A_{x}(t)=A_{x}^{mag}+\phi/L_{x}, \\
A_{y}(t)=A_{y}^{mag}+\theta/L_{y}
\eeq
The current operators $I_{x}$ and $I_y$ are given by
\beq
I_{x}=\frac{\partial \mathcal H}{\partial\phi}, I_{y}=\frac{\partial \mathcal H}{\partial\theta}
\eeq
One imposes the external Hall voltage ${\cal E}$ by varying the
parameter $\theta$ in time. In particular from
$\partial_{t}A_{y}=E_{y}$ and $-L_{y}E_{y}={\cal E}$ we see that
$\partial_{t}\theta={\cal E}$. Thus we can view the Hall voltage as the rate of change of the parameter $\theta$. Explicitly applying the expressions
for the transition amplitudes derived in the previous appendix, we
get that in the linear response
\beq
\langle I_{x}\rangle =i \sum_{m\neq 0}\frac{\langle
0|I_{x}|m\rangle\langle m|I_y|0\rangle- x\leftrightarrow y}{(\mathcal E_{m}-\mathcal E_{0})^2}\,\dot\theta.
\eeq
where $|0\rangle$ is an instantaneous ground state. The first term
in the RHS of this equation is exactly the curvature of the Berry
phase in the space of parameters $\theta$ and $\phi$:
\be
\mathcal F_{\theta\phi}=i\left(\langle 0|\overleftarrow{\partial_\theta}
\partial_\phi|0\rangle-\langle 0|\overleftarrow{\partial_\phi}\partial_\theta|0\rangle\right)
\ee

Averaging the Berry curvature over $\theta,\phi$ is equivalent to
the evaluation of the (first) Chern character:
\beq
ch_{1}(|0\rangle)=\frac{1}{2\pi}\int_{T^{2}}d\theta d\phi \mathcal F_{\theta\phi},
\eeq
which is an integer number.  This result gives the well known
topological quantization condition of the Hall conductivity:
\beq\label{chern-hall}
\sigma_{H}=\frac{e^{2}}{2\pi h}ch_{1}(|0\rangle)
\eeq

{\it Acknowledgments} We acknowledge useful discussions with C. Chamon, C. De\;\;Grandi, G.~Ortiz, A. Sandvik and M. Tomka. The work was supported by Grants NSF DMR-0907039 (AP), AFOSR FA9550-10-1-0110 (AP), the
Sloan Foundation (AP) and by Swiss National Science Foundation (VG).

Note added. After this work was complete we became aware of a recent paper~\cite{avron_11}, where partially overlapping results were obtained.




\end{document}